\documentclass[aps,preprint]{revtex4}
\usepackage{graphicx}
\usepackage{hyperref}
\usepackage{natbib}
\begin{document}
\begin{center}
\bfseries
\large
Quantum Impurity in a Nearly Critical Two Dimensional
Antiferromagnet\\
\normalsize
\mdseries
\vspace{0.5in}
Subir Sachdev${}^{\ast}$, Chiranjeeb Buragohain, and Matthias
Vojta
\end{center}
\begin{quote}
Department of Physics, Yale University, P.O. Box 208120, \\New Haven, CT~06520-8120,
USA\\
${}^{\ast}$ To whom correspondence should be addressed.
E-mail:
\href{http://pantheon.yale.edu/~subir}{subir.sachdev@yale.edu}
~\\
~\\
Published in \href{http://www.sciencemag.org}{{\sl Science}} {\bf 286}
2479 (1999).
\end{quote}
\newpage
\vspace*{2in}
\begin{quote}
We describe the spin dynamics
of an arbitrary localized impurity in an
insulating two dimensional antiferromagnet, across the
host transition from a paramagnet with a spin gap to a N\'{e}el
state. The impurity spin susceptibility
has a Curie-like divergence at the quantum-critical coupling, but
with a universal, effective spin which is neither an
integer nor a half-odd-integer.
In the N\'{e}el
state, the transverse impurity susceptibility is a universal number
divided by the host spin stiffness (which determines the energy cost to slow twists
in the orientation of the N\'{e}el order).
These, and numerous other results for the thermodynamics, Knight shift, and
magnon damping have significant applications to experiments
on layered transition metal oxides.
\end{quote}
\newpage

The recent growth in the study of quasi two dimensional
transition metal oxide compounds \emph{(\cite{rice})} with a paramagnetic
ground state and an energy gap to
all excitations with a non-zero spin (the `spin gap' compounds,
like ${\rm Sr Cu}_2 {\rm O}_3$, ${\rm Cu Ge O}_3$ and ${\rm Na
V}_2 {\rm O}_5$) has led to fundamental advances in our
understanding of low dimensional, strongly correlated electronic
systems. These systems are insulators and are so
not as complicated as the cuprate high
temperature superconductors (which display a plethora of phases
with competing magnetic, charge, and superconducting order), and
this simplicity has clearly exposed the novel characteristics of
the collective quantum spin dynamics.

One of the most elegant probes of these spin gap compounds is
their response to intentional doping by non-magnetic impurities,
like ${\rm Zn}$ or ${\rm Li}$, at the location of the magnetic
ions. Such experiments were initially undertaken on the cuprate
superconductors \emph{(\cite{fink,alloul})}, but their analogs in the
insulating spin gap compounds have proved to be a most fruitful
line of investigation \emph{(\cite{hase})}.
They have demonstrated a remarkable
property of the paramagnetic ground state of the host compound:
each non-magnetic impurity has a net magnetic moment of spin
$1/2$ located in its vicinity (for the case where the host
compound has magnetic ions with spin $1/2$). The confinement of
spin is a fundamental defining property of the host paramagnet,
and is a key characterization of the quantum coherent manner in
which the host spins form a many-body, spin zero ground state: this
confining property was predicted theoretically \emph{(\cite{fink,rs})} for
the paramagnetic states of a large class of two dimensional
antiferromagnets.

We will describe the quantum theory of an arbitrary
localized deformation in such antiferromagnets; examples of
deformations are ({\em i\/}) a single non-magnetic impurity,
along with changes in the values of nearby exchange interactions;
({\em ii\/}) change in sign of a localized group of exchange
interactions from antiferromagnetic to ferromagnetic.
Our main concern will be the
behavior of the impurity as the host antiferromagnet undergoes a
bulk quantum phase transition from a paramagnet to a magnetically
ordered N\'{e}el state:
we will show that the spin
dynamics of any deformation
is universally determined by a single number---an integer
or half-odd-integer valued spin $S$.

\begin{figure}[h!]
\centerline{\includegraphics[width=5in]{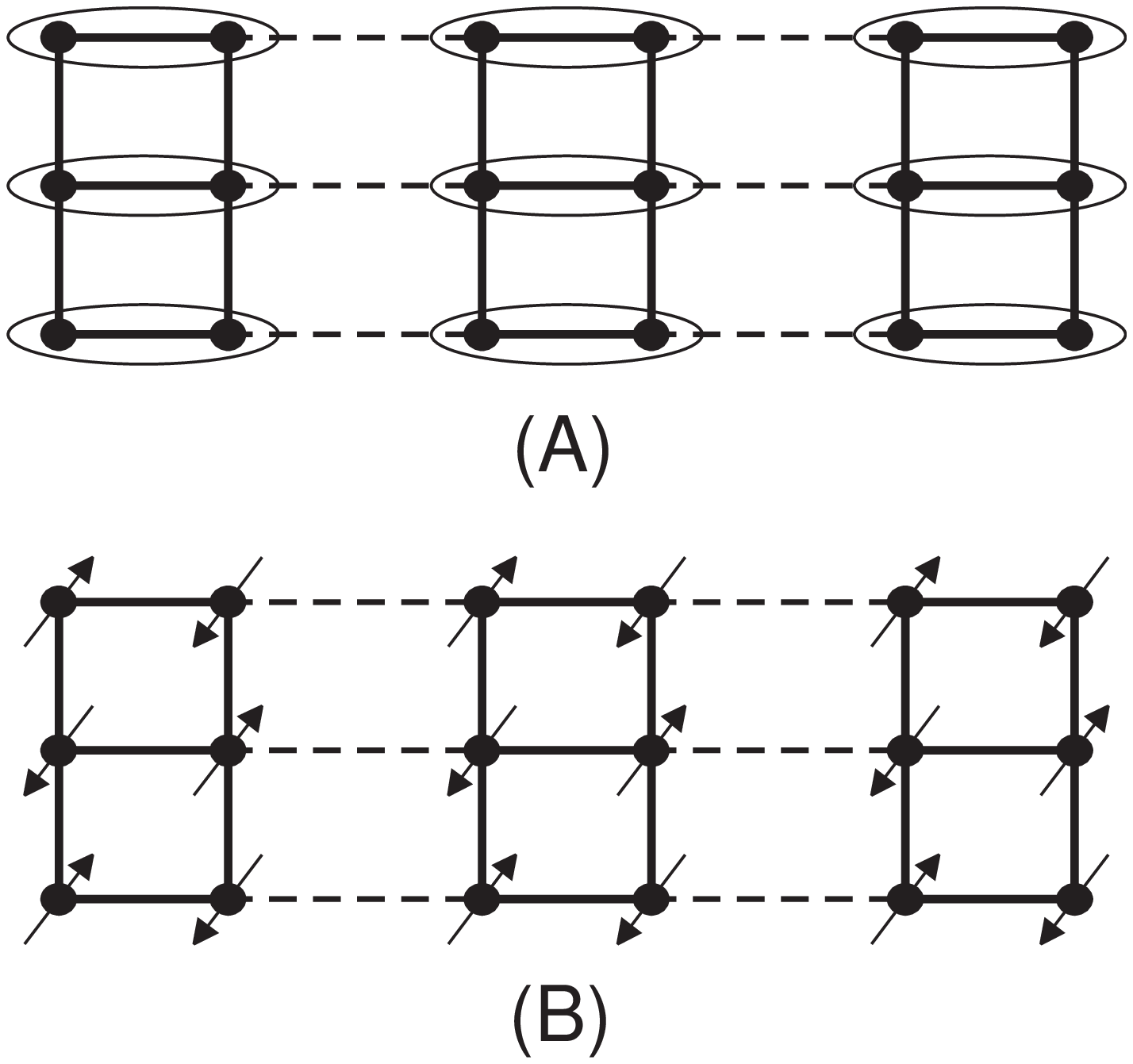}}
\vspace{0.3in}
\caption{The coupled ladder antiferromagnet. The $A$ links are full lines and
have exchange $J$, while the $B$ links are dashed lines and have exchange
$\lambda J$. The paramagnetic ground state for $\lambda < \lambda_c$
is schematically indicated in (A): the ellipse represents a
singlet valence bond,
$(|\uparrow \downarrow \rangle - |\downarrow \uparrow \rangle)/\protect\sqrt{2}$
between the spins on the sites. The N\'{e}el ground state for $\lambda > \lambda_c$
appears in (B).
}
\label{fig1}
\end{figure}

Apart from applications to experiments on materials intentionally
driven across a quantum phase transition,
our results also lead to new insights and predictions for
the behavior of impurities in existing spin gap compounds.
The traditional view of the
spin gap paramagnet is based on strong local singlet
formation between nearest-neighbor spins (see Fig~\ref{fig1}A
below); the resulting picture of doping by a
non-magnetic impurity is that the partner spin of the impurity
site is essentially free. To obtain any non-trivial dynamics one
performs an expansion about such a decoupled limit,
and this yields simple localized spin behavior with non-universal
details depending upon the specific microscopic couplings.
In practice, however, spin gap systems are usually well away from
the local singlet regime, and strong resonance between different
singlet pairings leads to appreciable spin correlation lengths: their
spin gap, $\Delta$, is significantly smaller than $J$,
a typical nearest-neighbor exchange.
A systematic and controlled approach for
analyzing such a fluctuating singlet state, which we advocate here,
is to find a quantum critical point to a magnetically ordered
state somewhere in parameter space, and to then
expand away from it into the
spin gap state. As we shall discuss below,
the coupling between the bulk and impurity excitations
becomes {\em universal} in such an expansion, and all dynamical
properties depend only upon the bulk parameters, $\Delta$ and a
velocity $c$ (defined below).

For clarity, we will state our main results in the context of a
simple, explicit theoretical model; however, they are
more general, and apply quantitatively to a broad class of
experimentally realizable systems. We begin by reviewing the
properties of the regular antiferromagnet described by the
Hamiltonian \emph{(\cite{katoh,book})}
\begin{equation}
{\cal H} = J \sum_{i,j \in A} {\bf S}_i \cdot {\bf S}_j + \lambda
J \sum_{i,j \in B} {\bf S}_i \cdot {\bf S}_j
\label{ham}
\end{equation}
where ${\bf S}_i$ are spin-1/2 operators on the sites of the
coupled-ladder lattice shown in Fig~\ref{fig1}, with the $A$ links
forming two-leg ladders while the $B$ links couple the ladders.
The ground state of $H$ depends only on the dimensionless coupling
$\lambda$, and we will restrict our attention to $J>0$, $0 \leq
\lambda \leq 1$. At $\lambda=0$ the ladders are decoupled, and
each forms a spin singlet quantum paramagnet (Fig~\ref{fig1}A).
This paramagnetic state continues adiabatically
for small non-zero $\lambda$ until the quantum critical coupling
$\lambda = \lambda_c \approx 0.3$, where the spin gap vanishes
as $\Delta \sim (\lambda_c - \lambda)^{\nu}$, where $\nu$ is a known
exponent \emph{(\cite{book})} (the symbol $\sim$ indicates the two quantities are
asymptotically proportional).
For $\lambda > \lambda_c$, the ground state
has long range N\'{e}el order (Fig~\ref{fig1}B)
characterized by the non-zero spin
stiffnesses, $\rho_{sx}, \rho_{sy}$, which determine the energy
cost to twists in the order parameter orientation in the $x$, $y$
directions (we also define $\rho_s \equiv (\rho_{sx}
\rho_{sy})^{1/2}$). The low-lying excitations above the N\'{e}el
state are spin-waves which travel with velocities $c_x$, $c_y$ in
the $x$, $y$ directions (with $c_x^2/c_y^2 = \rho_{sx}/\rho_{sy}$;
we define $c \equiv (c_x c_y)^{1/2}$). As $\lambda$ approaches the
critical value $\lambda_c$ from above, all the
stiffnesses vanish as $(\lambda - \lambda_c)^{\nu}$, while the
velocities remain finite and non-critical.

\begin{figure}[ht]
\centerline{\includegraphics[width=5in]{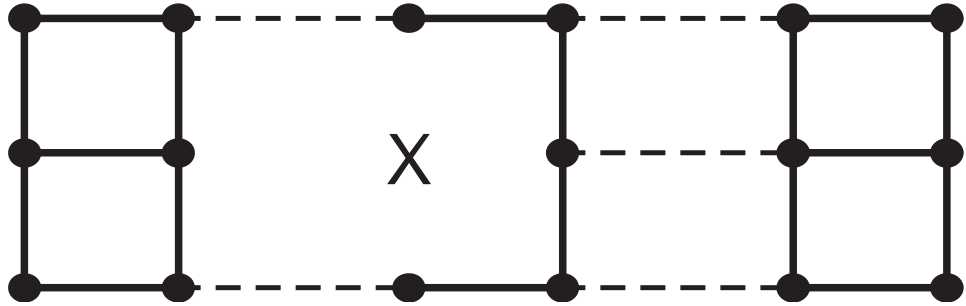}}
\vspace{0.3in}
\caption{The impurity Hamiltonian ${\cal H}_X$ in which the spin and links
on site $i=X$ have been removed.}
\label{fig2}
\end{figure}

Introducing a non-magnetic impurity in ${\cal H}$ by
removing the spin at site $i=X$ (Fig~\ref{fig2}),
the modified Hamiltonian ${\cal H}_X$ has the same form as ${\cal H}$
but all links connected to site $X$ do not appear in the sums in
Eq~\ref{ham}. The system can be probed by examining its total linear
susceptibility ($\chi$) to a uniform magnetic field, ${\bf H}$
(under which the Hamiltonian becomes $ {\cal H}_X - g \mu_B \sum_{i
\neq X} {\bf H} \cdot {\bf S}_i$ where $\mu_B$ is the Bohr
magneton and $g$ is the gyromagnetic ratio of the ion).
This susceptiblity may be written
as $\chi = (g \mu_B)^2 ({\cal A} \chi_b + \chi_{{\rm
imp}})$ where ${\cal A}$ is the total area of the antiferromagnet,
$\chi_b$ is the bulk response per unit area of the antiferromagnet
without the impurity, and $\chi_{{\rm imp}}$ is the additional
contribution due to the non-magnetic impurity. We will now
 describe the behaviors of $\chi_b$ and
$\chi_{\rm imp}$ as the temperature $T \rightarrow 0$, and
$\lambda$ moves across $\lambda_c$.

In the quantum paramagnet, $\lambda < \lambda_c$,
the presence of the spin gap implies
that the bulk response is exponentially small, $\chi_b =
(\Delta/\pi \hbar^2 c^2) e^{-\Delta/k_B T}$ \emph{(\cite{book})}.
The confinement of a magnetic
moment in the vicinity of the impurity site
implies that there will be Curie like contribution, and so
\begin{equation}
\chi_{{\rm imp}} = \frac{S(S+1)}{3 k_B T},
\label{chi1}
\end{equation}
where $S=1/2$ for the model under consideration here \emph{(\cite{sandvik})};
for a general local deformation, we
consider Eq~\ref{chi1} as the definition of the value of $S$, which,
naturally, must be an integer or a half-odd-integer.
These expressions for $\chi_b$ and $\chi_{{\rm imp}}$ are exact as
$T \rightarrow 0$ for all $0 < \lambda < \lambda_c$.
Another way of characterizing the confinement of the magnetic
moment near $X$ is by looking at the time autocorrelation function of a
spin at a site $i=Y$ close to $X$ (say, its nearest neighbor); at $T=0$
this obeys
\begin{equation}
\lim_{\tau \rightarrow \infty}
\left\langle {\bf S}_Y (\tau) \cdot {\bf S}_Y (0) \right\rangle =
m_Y^2 \neq 0,
\label{my}
\end{equation}
where $\tau$ is imaginary time, and
$m_Y$ is the local remnant magnetic moment on site $Y$, which
is usually significantly smaller than the total impurity moment $S$
appearing in Eq~\ref{chi1}.

Next, we turn to the behavior as $T \rightarrow 0$ at the critical point
$\lambda=\lambda_c$ (more generally, the $T>0$ results here will apply
for $\Delta < T < J$ ($\rho_s < T < J$) for $\lambda < \lambda_c$
($\lambda > \lambda_c$)). We expect that as the spin gap in the
quantum paramagnet disappears, the bulk magnon excitations will proliferate
and their screening will
eventually quench the impurity moment---so $m_Y$ approaches $0$ as
$\lambda$ approaches $\lambda_c$ from below.
We can anticipate a power-law decay of the spin
autocorrelations \emph{(\cite{sy,anirvan,si})}, with
\begin{equation}
\left\langle {\bf S}_Y (\tau) \cdot {\bf S}_Y (0) \right\rangle
\sim 1/\tau^{\eta^{\prime}}
\label{etai}
\end{equation}
for large $\tau$, $T=0$, and $\lambda=\lambda_c$,
and our result for the new, universal exponent $\eta^{\prime}$
is given below;
standard scaling arguments also imply that $m_Y$ vanishes
as $m_Y \sim (\lambda_c - \lambda)^{\eta^{\prime} \nu/2}$. The behavior at
the critical point therefore appears analogous
to that in the overscreened
multichannel Kondo problem \emph{(\cite{noz,afflud})};
in that case, the impurity spin is
screened by a bath of conduction electrons carrying multiple `flavors',
and also exhibits a power-law decay in its autocorrelation.
Furthermore, in the multichannel Kondo case, the $T$
dependence of $\chi_{{\rm imp}}$ is given essentially by the
Fourier transform of Eq~\ref{etai}, that is by $\chi_{{\rm imp}} \sim T^{-1 +
\eta^{\prime}}$ \emph{(\cite{afflud})}. This result is a consequence of a `compensation'
effect \emph{(\cite{pwa})}, as the magnetic response of the screening cloud of conduction
electrons is negligible: the local Fermi levels of up and down
electrons adjust themselves to the local magnetic field, and hence the
susceptibility is not very different from the bulk susceptibility
except in the immediate vicinity of the impurity spin \emph{(\cite{barzykin})}. In more
technical terms, $\chi_{{\rm imp}}$ vanishes in the strict
continuum limit, and corrections to scaling have to be considered,
which lead eventually to $\chi_{{\rm imp}} \sim T^{-1 +
\eta^{\prime}}$. Our computations show that the behavior of ${\cal H}_X$
at $\lambda=\lambda_c$ is dramatically different: the magnon
excitations do not have an exact compensation property, and
their response is non-zero already in the scaling limit. So in a sense,
the present problem is simpler than the overscreened Kondo case,
and naive scaling arguments always work, without inclusion of
irrelevant operators---the scaling dimension of
$\chi$ is that of inverse energy \emph{(\cite{book})}, and so we have one
of our central results:
\begin{equation}
\chi_{{\rm imp}} = \frac{{\cal C}_1}{k_B T}
\label{chi2}
\end{equation}
at $\lambda=\lambda_c$, where ${\cal C}_1$ is a
universal number independent of microscopic details
(as are all the ${\cal C}_i$ introduced below).
We computed ${\cal C}_1$ in the expansion in $\epsilon=3-d$, where $d$ is the
spatial dimension, and obtained
\begin{equation}
{\cal C}_1 = \frac{S(S+1)}{3} \left[ 1 + \left(\frac{33 \epsilon}{40}\right)^{1/2}
 -\frac{7 \epsilon}{4} + {\cal O} \left(
\epsilon^{3/2} \right) \right];
\label{chi3}
\end{equation}
the omitted higher order corrections in Eq~\ref{chi3} will, in
general, depend upon $S$.
Comparing with Eq~\ref{chi1} we can define an effective
impurity spin, $S_{\rm eff}$, at the quantum-critical point
by ${\cal C}_1 = S_{\rm eff} ( S_{\rm eff} + 1)/3$;
it is evident that $S_{\rm eff}$ is a universal function of $S$,
is neither an integer
nor a half-odd-integer, and is almost certainly irrational at $\epsilon=1$.
 Also notice that the leading corrections
in the $\epsilon$-expansion are quite large, and this will be a
feature of all the results obtained below; accurate numerical
estimates require some resummation scheme, but we will
not discuss this here.
For completeness, let us note that at $\lambda=\lambda_c$,
the bulk response \emph{(\cite{CS})}
$\chi_b = {\cal C}_2 (k_B T)/(\hbar c)^2 $, a $T$-dependence that
is also different from the bulk response in the overscreened Kondo
problem.

\begin{figure}[ht]
\centerline{\includegraphics[width=7.5in]{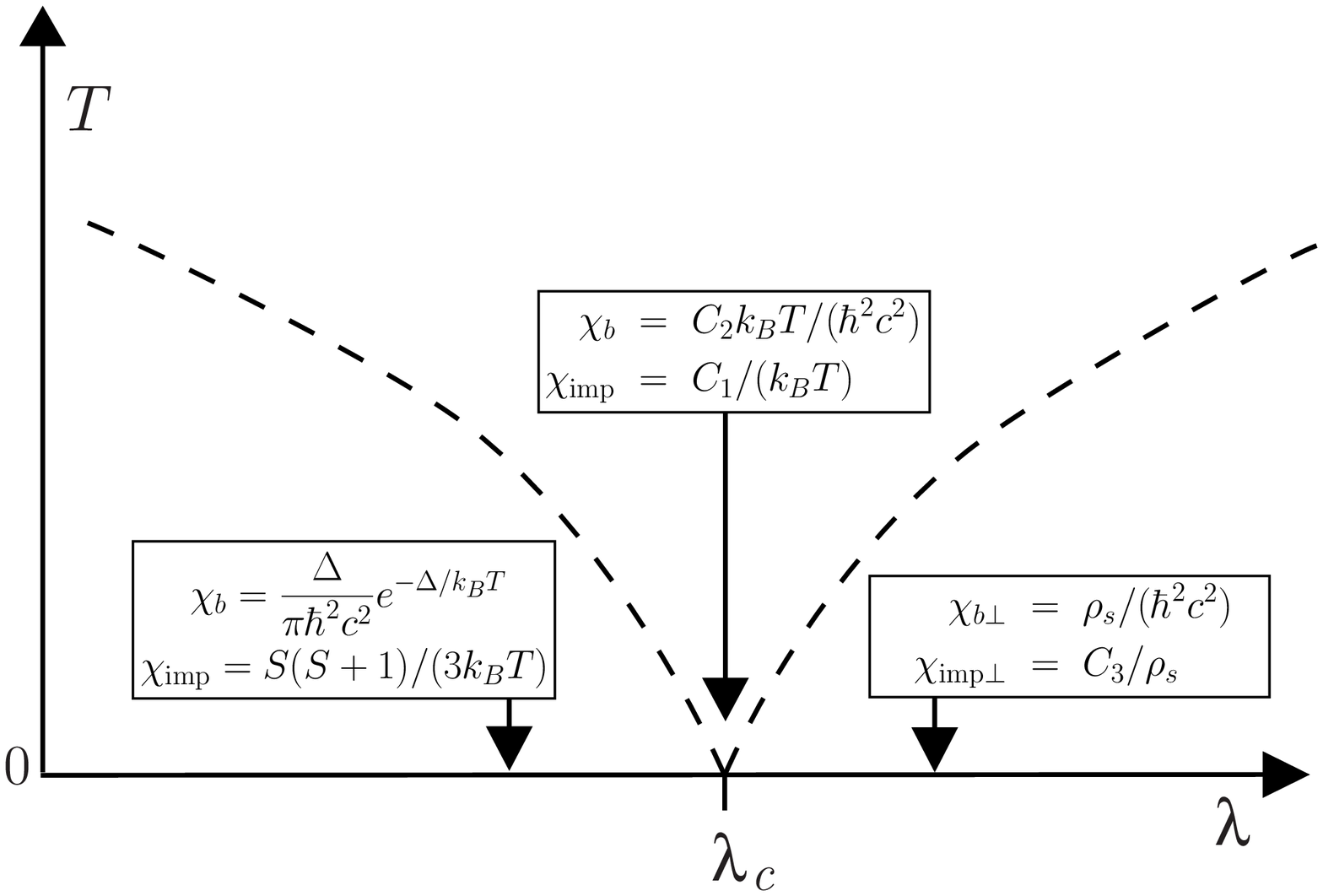}}
\caption{Summary of the results for the bulk and impurity susceptibilities of
${\cal H}_X$. The
constants ${\cal C}_{1-3}$ are universal numbers, insensitive to microscopic
details like
variations in the magnitude or sign of the exchange constants in the vicinity of the
impurity, or presence of additional, nearby, vacancies or impurity ions
with different spins. The constants ${\cal C}_1$ and ${\cal C}_3$
depend only on the integer/half-odd-integer valued $S$, and
we can view Eq~\protect\ref{chi1}, the $T\rightarrow 0$ limit of
$\chi_{\rm imp}$ in the paramagnet ($\lambda < \lambda_c$),
as the experimental definition of $S$. For the
case in which non-magnetic impurities are added in a localized
region, with no modification of exchange constants, $S$ is the net
imbalance of spin between the two sublattices.
The constant ${\cal C}_1$ defines the effective spin at the
quantum-critical point
by ${\cal C}_1 = S_{\rm eff} (S_{\rm eff} + 1)/3$, and $S_{\rm eff}$ is
neither an integer nor a half-odd-integer.
}
\label{fig3}
\end{figure}

Finally, we describe the situation for $\lambda > \lambda_c$. Now
the presence of N\'{e}el order at $T=0$ implies that the response is
anisotropic. Parallel to the N\'{e}el order,
there is a total magnetic moment quantized precisely at $S$ \emph{(\cite{sandvik})},
and this does not vary
under a small longitudinal field (there is also a staggered
local moment in zero field,
as defined by Eq~\ref{my}, which obeys
$m_Y \sim
|\lambda-\lambda_c|^{\eta^{\prime}\nu/2}$).
Orthogonal to the N\'{e}el order, there is
a linear response to a transverse field, $\chi_{\perp}$.
For the bulk response, we have
the well-known result that $\chi_{b \perp}$ is proportional to the
spin stiffness, $\chi_{b \perp} = \rho_s /(\hbar c)^2$. In
contrast,
the same scaling arguments leading to Eq~\ref{chi2} imply that $\chi_{{\rm imp}\perp}$
is inversely proportional to $\rho_s$, the latter being the
only energy scale characterizing the ground state as $\lambda$
approaches
$\lambda_c$ from above; so another key result of this paper is
\begin{equation}
\chi_{{\rm imp} \perp} = \frac{{\cal C}_3}{\rho_s}.
\label{chi4}
\end{equation}
In general $d$, this relationship is $\chi_{{\rm imp} \perp} = {\cal C}_3
(\hbar c)^{(2-d)/(d-1)}/\rho_s^{1/(d-1)}$, and the $\epsilon$-expansion of
${\cal C}_3$ is
\begin{equation}
{\cal C}_3 = \frac{15 S}{\sqrt{22}} \left( \frac{11 S_{d+1}}{2 \epsilon}
\right)^{1/(d-1)} \left[1 -(1.193 + 0.553 S + 0.419 S^2) \epsilon +
{\cal O}(\epsilon^2)\right],
\label{chi5}
\end{equation}
where $S_d = 2/(\Gamma(d/2) (4
\pi)^{d/2})$.
Note that
$\rho_s$ vanishes, and so $\chi_{{\rm imp} \perp}$ diverges, as
$\lambda$ approaches
$\lambda_c$. Turning to $T>0$ but very small, in $d=2$ and in the absence of any
spin anisotropy, strong angular fluctuations cause
the N\'{e}el order to vanish at any non-zero $T$.
Then the susceptibility takes the rotationally averaged value $\chi_{\rm imp}=
S^2 /(3 k_B T) + (2/3) \chi_{{\rm imp} \perp}$, where the first term is the
contribution of the net moment noted earlier (note that this term has
a coefficient $S^2$ and not $S(S+1)$, because the locking of the
moment orientation
to the local N\'{e}el order makes it behave classically). In practice, this
averaged $\chi_{\rm imp}$ will not be observable as even an extremely small
anisotropy will pin the N\'{e}el order below a small $T>0$.
Our results for $\chi$ are summarized in
Fig~\ref{fig3}.

The next two paragraphs contain a technical interlude which outlines the
field-theoretic derivation of the results above---details appear
elsewhere \emph{(\cite{long})}. We describe the bulk-ordering transition by
a $d+1$-dimensional field theory with action ${\cal S}_b$
of a field $\phi_{\alpha} (x, \tau)$
($\alpha = 1\ldots3$) representing the collinear N\'{e}el order
parameter \emph{(\cite{book})}. This is coupled by the action ${\cal S}_{\rm imp}$
to an impurity spin at $x=0$
with
orientation given by the unit vector $n_{\alpha}$. The partition
function is
$\int {\cal D} \phi (x, \tau) {\cal D} n(\tau) \exp (-{\cal S}_b - {\cal S}_{\rm imp})$
with
\begin{eqnarray}
{\cal S}_b = && \int d^d x d \tau \left[ \frac{1}{2} \left( (\nabla_x
\phi_{\alpha})^2 + c^2 (\partial_\tau \phi_{\alpha})^2 + r
\phi_{\alpha}^2 \right) + \frac{g_0}{4!} (\phi_{\alpha}^2)^2 \right]
\label{actionb} \\
{\cal S}_{\rm imp} = && \int d \tau \left[ i S A_{\alpha} (n)
\frac{d n_{\alpha}}{d \tau} - \gamma_0 S n_{\alpha}( \tau ) \phi_{\alpha} (x=0,\tau)
\right]
\label{actionimp}
\end{eqnarray}
where $\epsilon_{\alpha\beta\gamma} \partial A_{\beta} /\partial n_{\gamma} =
n_{\alpha}$, and the term proportional to $A(n)$ is a Wess-Zumino
form representing the Berry phase of
the impurity spin. The bulk transition in ${\cal S}_b$ is driven
by tuning the coupling $r$ through a critical value $r_c$, which
therefore plays a role similar to $\lambda$; the $\lambda <
\lambda_c$ ($\lambda > \lambda_c$)
region of the lattice antiferromagnet ${\cal H}$ maps onto the
$r>r_c$ ($r<r_c$) region of the field theory ${\cal S}_b$.
Quite generally, any local deformation of the
antiferromagnet is
described by the action ${\cal S}_b + {\cal S}_{\rm imp}$,
where $S$, defined as the integer or
half-odd-integer appearing in Eq~\ref{chi1}, is (roughly)
the net local imbalance of
spin between the two sublattices.
Changes in
exchange constants lead to additional terms
like $\int d \tau \phi_{\alpha} ^2 ( x=0, \tau)$ which are all
strongly irrelevant under the renormalization group (RG) analysis in powers
of $\epsilon$. The $r=0$, $g_0=0$ case of Eqs~\ref{actionb},\ref{actionimp} was
considered earlier by Sengupta \emph{(\cite{anirvan})} (and related models
in \emph{(\cite{sy,si})}) in a non-local
formulation in which $\phi_{\alpha} (x \neq 0, \tau)$ was
integrated out: however, such a model has a pathological response
to even an infinitesimal field ${\bf H}$ (the energy is
unbounded below), and the quartic $g_0$ coupling is essential to stabilize
the system, and to all the
results obtained here. Further, the local formulation here
facilitates
development of the RG to all orders.

The RG analysis of ${\cal S}_b+{\cal S}_{\rm imp}$ is carried out by the methods of
`boundary critical phenomena' \emph{(\cite{cardybook})} of a $(d+1)$-dimensional system with a
1-dimensional `boundary' at $x=0$, which constitutes a `dimensional reduction' of
$d>1$ (contrast this with the case of a
$(d+1)$-dimensional system with a $d$-dimensional boundary, with a dimensional
reduction of 1, which has
been invariably \emph{(\cite{afflud,diehl})} considered earlier, as in all the
Kondo problems). The irrelevance of the boundary `mass' term
$\phi_{\alpha} ^2 ( x=0, \tau)$ implies that there is only an
`ordinary' transition at the position of the bulk critical
point \emph{(\cite{diehl2})} (this has been implicit in our earlier discussion),
and there are no analogs of the `surface', `special', and
`extraordinary' transitions \emph{(\cite{cardybook})}.
The RG analysis of the bulk action ${\cal S}_b$ is now standard
text-book material---we will not reproduce it here, and will
follow the notation of \emph{(\cite{bgz})}. We introduce renormalized
fields $\phi = \sqrt{Z} \phi_R$, $n = \sqrt{Z^{\prime}} n_R$, and
renormalized couplings by $g_0 = (\mu^{\epsilon}/c) (Z_4/Z^2 S_{d+1})
g$, $\gamma_0 = (\mu^{\epsilon} c)^{1/2} (Z_{\gamma}/\sqrt{Z Z^{\prime}
\widetilde{S}_{d+1}}) \gamma$
where $\mu$ is a renormalization
inverse length scale, $\widetilde{S}_d = \Gamma(d/2-1)/(4 \pi^{d/2})$, and
the bulk renormalization factors $Z$, $Z_4$ are specified in
\emph{(\cite{bgz})}. For the new boundary renormalization factors, we
obtained to two loops $Z^{\prime} = 1 - 2 \gamma^2 /\epsilon + \gamma^4 /\epsilon$
and $Z_{\gamma} = 1 + \pi^2 (S(S+1)-1/3) g \gamma^2/(6 \epsilon)$.
These lead to the $\beta$ function for $g$ found in \emph{(\cite{bgz})},
and the new $\beta$ function for the boundary coupling
\begin{equation}
\beta ( \gamma) = - \frac{\epsilon \gamma}{2} + \gamma^3 -
\gamma^5 + \frac{5 g^2 \gamma}{144} + \frac{\pi^2}{3} (S(S+1)-1/3) g
\gamma^3 + {\cal O} \left((\gamma, \sqrt{g})^7 \right)
\end{equation}
The critical fluctuations at the boundary are therefore controlled
by the fixed point values $\gamma = \gamma^{\ast}$,
$g = g^{\ast}$ (both nonzero)
at which both $\beta$ functions vanish, and
canonical methods then imply the exponent
\begin{equation}
\eta^{\prime} = \epsilon - \left[ \frac{5}{242} + \frac{2
\pi^2}{11} \left( S(S+1)-1/3 \right) \right] \epsilon^2 + {\cal O}
(\epsilon^3).
\end{equation}
Eq~\ref{chi3} can now be obtained by the methods of
\emph{(\cite{ss})}, while Eq~\ref{chi5} follows directly from a
renormalized perturbation theory in the ordered phase at $T=0$.
We conclude our technical interlude by noting that our RG scheme shows directly
that the only graphs which contribute to the renormalization of
$\gamma_0$, beyond those arising from the wavefunction
renormalization $Z^{\prime}$, must include a factor of the bulk
interaction $g_0$; this implies that
$Z_{\gamma} = 1$ for $g=0$, and
shows that for the models of \emph{(\cite{sy,anirvan})}
the one-loop exponent $\eta^{\prime} = \epsilon$ is exact.

The above methods can be extended to determine the behavior of
other observables in the regimes of Fig~\ref{fig3}. We mention a
few:
\newline
({\em i\/}) Entropy: In the paramagnetic phase ($\lambda <
\lambda_c$) there is clearly a residual entropy of $\ln
(2S+1)$ as $T \rightarrow 0$. At $\lambda=\lambda_c$, the
$\epsilon$-expansion shows that this is modified to $\ln(2S+1) -
S(S+1)(33 \epsilon/160)^{1/2} + {\cal O}(\epsilon^{3/2})$, while in the
N\'{e}el state ($\lambda > \lambda_c$, the N\'{e}el order pinned
by some small spin anisotropy) the impurity entropy vanishes as $T^{d}$
at low $T$.
\newline
({\em ii\/}) Knight shift: We restrict the discussion here to the
intermediate quantum-critical region of Fig~\ref{fig3}, $T > |\lambda -
\lambda_c|^{\nu}$. The shift in the NMR resonance frequency is
proportional to the local response in the presence of a
uniform external field, $\chi (x)$. In the vicinity of the impurity ({\em
e.g.} at site $i=Y$)
$\chi(x) \sim T^{-1+\eta^{\prime}/2}$. Well away from the impurity
($|x| \rightarrow \infty$), apart from the bulk response of the
antiferromagnet, there are staggered and uniform contributions
which decay exponentially with $|x|$ on a scale $\sim \hbar c/(\sqrt{\epsilon} k_B
T)$.
\newline
({\em iii\/}) Magnon damping: In the quantum paramagnet ($\lambda <
\lambda_c$), and at $T=0$,
the pure antiferromagnet has a pole in the dynamic spin structure factor
$\sim 1/(\Delta - \hbar \omega)$ at the antiferromagnetic ordering wavevector
from the triplet magnon
excitations. In the presence of a dilute
concentration of impurities, $n_i$,
this pole will be broadened on an energy scale $\Gamma$;
scaling arguments and
the structure of the fixed
point found here imply the exact form \emph{(\cite{precise})} $\Gamma \sim n_i
(\hbar c)^{d}
\Delta^{1-d}$. We argue that this damping mechanism is the
main ingredient in the broadening of the `resonance peak'
observed recently in ${\rm Zn}$-doped ${\rm Y Ba}_2 {\rm Cu}_3
{\rm O}_7$ \emph{(\cite{keimer})}. Using the values $\hbar c = 0.2 a$ eV
($a$ is the lattice spacing),
$\Delta = 40$ meV, and $n_i = 0.005/a^2$, we obtain the estimate
$\Gamma = 5$ meV, which is in excellent accord with the observed
linewidth of 4.25 meV \emph{(\cite{keimer})}. We have also studied the lineshape
of the magnon peak \emph{(\cite{long})}, and find that it is
asymmetric at very low $T$,
with a tail at high frequencies:
it would be interesting to test this in future experiments.

We have described the highly non-trivial, collective,
quantum spin dynamics of a single impurity in a strongly
correlated, low dimensional electronic system. The
problem maps onto a new boundary quantum
field theory, Eqs~\ref{actionb},\ref{actionimp}, and is therefore also
of intrinsic theoretical
interest: unlike previously studied quantum impurity problems,
there is a
complicated interference between bulk and boundary interactions,
and its proper description is the key to the physical results we
have obtained.
Our theoretical results for the magnon damping in the spin gap
phase are in good agreement with existing
experiments \emph{(\cite{keimer})}. Studies of materials
exhibiting other aspects of the regimes of Fig~\ref{fig3}
appear possible, and we hope they will be undertaken;
spin gap compounds can be driven across the
transition by, say, application of hydrostatic pressure, or by
doping with other impurities which have the same spin as the host
ion they replace and do not change the sign of the exchange
constants \emph{(\cite{renard})}.
Quantum Monte Carlo simulations should also allow
more accurate determination of the universal constants ${\cal
C}_1$ and ${\cal C}_3$.

\newpage
\section*{Figures}
\renewcommand{\baselinestretch}{1.5}
\vspace{0.3in}

\newpage

\newpage

\begin{thebibliography}{}
\bibitem{precise} More precisely, the theory
in Eqs~\ref{actionb},\ref{actionimp} predicts that $\Gamma = {\cal C}_4 n_i
(\hbar c)^{d} \Delta^{1-d}$ where ${\cal C}_4$ is a universal
number, of order unity in $d=2$. Note that this result is {\em independent}
of the coupling, $\gamma_0$, between the bulk and the impurity.
\bibitem{keimer} H.~F.~Fong et al., {\em Phys. Rev. Lett.} {\bf
82}, 1939 (1999).
As suggested by these authors, we are treating the superconductor
as a spin gap insulator: this is a good approximation because
the resonance excitation does not couple efficiently to the
superconducting quasiparticles.
\end{thebibliography}
\end{document}